# GENEOnet: A new machine learning paradigm based on Group Equivariant Non-Expansive Operators. An application to protein pocket detection.

**Giovanni Bocchi** [*,1], **Patrizio Frosini** [2], **Alessandra Micheletti** [1], **Alessandro Pedretti** [3]
**Carmen Gratteri** [4], **Filippo Lunghini** [5], **Andrea Rosario Beccari** [5] and **Carmine Talarico** [5]

[1] Department of Environmental Science and Policy, Università degli Studi di Milano, via Saldini 50, 20133 Milano, Italy

[2] Department of Mathematics, Università degli Studi di Bologna, Piazza di Porta S.Donato 5, 40126 Bologna, Italy

[3] Department of Pharmaceutical Sciences, Università degli Studi di Milano, via Mangiagalli 25, 20133 Milano, Italy

[4] Dipartimento di Scienze della Salute, Università degli Studi "Magna Græcia di Catanzaro", Campus " Salvatore Venuta", Viale Europa, 88100, Catanzaro, Italy.

[5] Dompé Farmaceutici SpA, Via Tommaso de Amicis 95, Napoli 80123, Italy

[*] giovanni.bocchi1@unimi.it

## Abstract

Nowadays there is a big spotlight cast on the development of techniques of explainable machine learning. Here we introduce a new computational paradigm based on Group Equivariant Non-Expansive Operators, that can be regarded as the product of a rising mathematical theory of information-processing observers. This approach, that can be adjusted to different situations, may have many advantages over other common tools, like Neural Networks, such as: knowledge injection and information engineering, selection of relevant features, small number of parameters and higher transparency. We chose to test our method, called GENEOnet, on a key problem in drug design: detecting pockets on the surface of proteins that can host ligands. Experimental results confirmed that our method works well even with a quite small training set, providing thus a great computational advantage, while the final comparison with other state-of-the-art methods shows that GENEOnet provides better or comparable results in terms of accuracy.

## Main

The recent advances in deep learning are revealing the growing importance of equivariant operators, mainly because of the need to make neural networks more transparent and interpretable[1,2,3,4,5,6,7,8] and to simplify the training phase. The use of such operators corresponds to the rising interest in the so called "explainable machine learning" [9,10,11], which looks for methods and techniques that can be understood by humans. In accordance with this line of research, group equivariant non-expansive operators (GENEOs) have been recently proposed as elementary components for building new kinds of neural networks[12]. Their use is grounded in Topological Data Analysis (TDA) and guarantees good mathematical properties, such as compactness, convexity, and finite approximability, under suitable assumptions on the space of data and with respect to the choice of appropriate topologies. Furthermore,

GENEOs allow to shift the attention from the data to the observers who process them, and to incorporate the properties of invariance and simplification that characterize those observers. The basic idea is that we are not usually interested in data, but in approximating the experts' behaviour in the presence of given data[13].

These properties open the way to a new kind of "geometric knowledge engineering for deep learning", which can allow us to drastically reduce the number of involved parameters and to increase the transparency of neural networks, by inserting information in the agents that are responsible for processing the data.

In this paper, we test this new paradigm by checking the advantages of using GENEOs for pocket detection in the context of drug development. Our purpose is to show that the theory of GENEOs can be practically and efficiently used in relevant applications, where other machine learning techniques have already been applied in a rather "blind way", without fully exploiting the a priori knowledge on the problem.

**Background on protein pocket detection**

The recent advances in computational technology have progressively increased the role of the *in silico* techniques in the drug discovery and repositioning processes[14]. The development of novel potential drugs can be carried out by different computational approaches[15]. Among them, the structure-based drug design[16] are attracting great interest due to their capacity to predict the binding affinity of novel compounds, to reveal the interacting substructures[17,18] and to elucidate the mechanism of action[19]. In this context, molecular docking techniques allow rapid and effective simulations of the molecular recognition process at an atomic level. To enhance their efficiency, docking calculations require a precise characterization of the binding pocket on which the search must be focused[20]. Blind docking analyses, which involve the whole target protein, lose efficiency in finding the correct ligand poses[21]. The putative binding sites can be clearly defined when co-crystallized ligand-receptor complexes and/or site-directed mutagenesis studies are available[22,23]. Without these data, the binding site cannot be precisely characterized even though the protein structure is experimentally resolved. Moreover, there may be the need to identify allosteric/accessory binding sites and prioritize the detected cavities according to their druggability[24]. Indeed, a reliable evaluation of the binding site druggability is a crucial requisite to optimize virtual screening campaigns involving large libraries of compounds on targets whose binding sites are unknown. Therefore, in the last years considerable efforts were focused on the development of algorithms and programs for identifying and characterizing potential binding sites. From a conceptual point of view, these tools involve two major tasks: a geometrical analysis for detecting the empty regions within a protein structure and a physicochemical analysis to characterize the interacting and structural properties of the found pockets in order to prioritize them and to identify the correct binding site(s).

Concerning the geometrical detection of the binding sites, the algorithms can be roughly subdivided into two main groups: grid-based and grid-free methods. The former[25-31] include various algorithms which calculate 3D grids and search which points are inside or outside the protein structure and satisfy some geometric and physicochemical conditions. POCKET[25] and its variants (LigSite[26], LigSite$^{csc}$ [27]) scan the grid to find patterns of protein-solvent-protein points to identify empty regions surrounded by the protein atoms. Several clustering approaches were developed to merge groups of points to a given cavity such as the methods based on buriedness index in Pocket-Picker[28] or on the calculation of potential maps[29]. SiteMap[30] characterizes the pockets by combining the local properties calculated for each grid point with the global properties resulting from the entire set of points. CAVIAR[31] further enhances this method with two main strengths: 1) the calculation of descriptors to predict the "ligandability" of a given binding site; 2) the capability to split the pockets into subcavities which can be used as pharmacophoric points. A grid of voxels can be used to efficiently represent volumes as in CavVis[32], which exploits them to detect the cavities in combination with the analyses of the Gaussian surfaces to approximate the solvent-excluded surface.

The grid-free approaches[33-38] use spherical probes, which are placed on the protein surface and clustered according to their properties which are representative of candidate pockets. SURFNET[33] searches for empty areas by positioning spheres between pairs of atoms and by reducing their radii until they do not intersect any atom. PASS[34] uses small spherical probes to coat the protein that are placed according to unique triplets of atoms, while the pockets are identified by counting the protein atoms within a distance from the probe to exclude the convex parts of the surface. Other grid-free approaches use the Voronoi tessellation to detect the cavities such CAST[35] and APROPOS[36], while SiteFinder[37] and Fpocket[38] are based on the concept of alpha spheres to detect local curvatures on the protein surface.

Regarding the algorithms proposed to prioritize the detected cavities, the probe-based methods use small molecules as representative probes across the surface to evaluate the interacting capacities of the binding sites. FTSite[39], which is a grid-based approach, scans the grid with 16 different probes for evaluating the interaction energy of each point by empirical energy functions. Q-SiteFinder[40] uses methyl probes and van der Waals energy to evaluate binding sites, while SILCS[41] uses MD simulations of the protein with fragment probes to build probability maps of binding.

Due to the recent advancements of data-intensive predictive approaches, the application of machine learning and deep learning techniques in the binding site prioritization comes as no surprise. P2Rank[42,43] uses a conventional machine learning algorithm (i.e. random forest) to evaluate the capability of each pocket point to bind a ligand. Deep learning techniques which are particularly effective in detecting complex relations between data, proved successful in binding site analyses. DeeplyTough[44] explores the potential binding sites by using a convolutional neural network (CNN). DeepSite[45] and DeepPocket[46] can be considered an evolution of P2Rank because they use a CNN to score the points on the protein surface and to cluster them for analyzing the binding pockets.

In this paper, we implement a model based on GENEOs, called GENEOnet, for protein pocket detection and scoring, and we compare its accuracy with a set of state-of-the-art methods described above.

Our results suggest that the approach to deep learning based on GENEOs can indeed offer many interesting benefits.

**Mathematical background on Group Equivariant Non Expansive Operators**

GENEOnet is a tool that blends the geometric and explainability properties of GENEOs with a network architecture into a novel knowledge-based machine learning paradigm.

GENEOs are mathematical tools with the aim of representing observers who analyze data. Observers can often be seen as functional operators, transforming data into other data, when data are represented by functions. This happens, for example, when we blur an image by a convolution. However, observers are far from being entities that merely change functions into other functions. They do that in a compatible way with respect to some group of transformations, i.e., they commute with these transformations. For example, the operator associating each regular function $f: \mathbb{R}^n \to \mathbb{R}$ to its Laplacian $\Delta f$ commutes with each Euclidean isometry of $\mathbb{R}^n$.

More precisely, we say that this operator is *equivariant* with respect to the group of isometries.

Observers are often additionally endowed with other regularity properties, in particular with *non-expansivity*. An operator representing an observer is non-expansive if the distance between the input data is not smaller than the distance between the output functions. This type of regularity is frequently found in applications, since in several cases our operators are required to simplify the metric structure of data. We can obviously imagine particular applications where this condition is locally violated, but the usual final purpose of data processing is to converge to an *interpretation*, i.e., a representation that is much simpler and meaningful than the original data.

As a consequence, it is reasonable to assume that the operators representing observers (or at least the composition of long enough sequences of such operators) are non-expansive. This assumption is not only useful to simplify the data analysis, but it is also fundamental in the proof that the space of group equivariant non-expansive operators is compact (and hence finitely approximable), provided that the space of data is compact with respect to a suitable topology[12].

See the Methods section for the formal definition of a GENEO.

## Results

GENEOnet exploits the rationale behind GENEOs to solve the problem of protein pocket detection. The paradigm has been here declined for the specific application here considered, but, with the right adjustments, it could be extended and applied to many other situations. Moreover, even though in this paper we will study an application of GENEOs, we remark that there is a mathematical theory describing the properties of GENEOs and their space. The interested reader will find more details in [12,47].

The main reason for choosing this specific application is due to some characteristics that make it very suitable to be treated with GENEOs. First of all there is some important empirical knowledge that is hard to embed in the usual machine learning techniques, but can easily be exploited by a method based on GENEOs.

For example it is known that binding sites tend to be in the lipophilic areas of the protein, otherwise they would continuously be filled with solvent, having thus no chances to interact with any other ligand. Another empirical rule says that if a pocket wants to host a binding site then it should be able to accept and donate some hydrogen bonds otherwise no ligand could find stable housing into that pocket.

Secondly if we rotate or translate a protein, its pockets will be as well transformed in the same way, coherently with the entire protein. This clearly implies that pocket detection is equivariant with respect to the group of spatial isometries.

We used these and some other pieces of information to design a pool of GENEOs able to identify promising binding sites. We first discretized the space surrounding the protein into voxels. Thus, GENEOnet falls in the group of grid-based computational methods. Our choice for the GENEOs fell on convolutional operators that process a set of "channels", i.e., functions that reflect a reasoned selection of geometric, physical, and chemical properties of a protein. The convolutional kernels of these operators have been designed with a knowledge engineering process to exploit all the information about the problem. Therefore, the final pool of GENEOs is composed by families of operators, each parametrized by a shape parameter. These parameters directly influence the shape of the kernels of the operators belonging to the corresponding family and thus the action of each single operator.

These families are then networked through a convex combination that allows us to explore a larger region of the space of GENEOs. Indeed these second-level operators depend on all the shape parameters and the convex combination weights. A last parameter is needed to transform the output of the method pipeline into a binary function. This function assigns to each voxel of the space surrounding the protein the value 1 if it belongs to a pocket and 0 otherwise.

The parameters of the model are then identified during an optimization step, that employs the Adam optimizer. Optimization is aimed to maximize our accuracy function that expresses a weighted ratio of correct recognition of voxels lying either inside pockets or outside pockets. Since GENEOnet depends eventually on 17 unknown parameters only, the optimization is performed using a small training set of ligand-protein complexes, that proved to be sufficient to obtain a quite good accuracy in pocket identification. As a byproduct of our model we also obtained a *druggability score* for each identified pocket. In this way it is possible to rank the pockets on the same molecule by scoring them in decreasing order. As a consequence, a set of models trained on different training sets of the same (small) size have been compared on a validation set, and the model providing the best accuracy in the pocket ranking has been selected.

The full training procedure and more details about the construction of the model can be found in the Methods section and in the Supplementary Information.

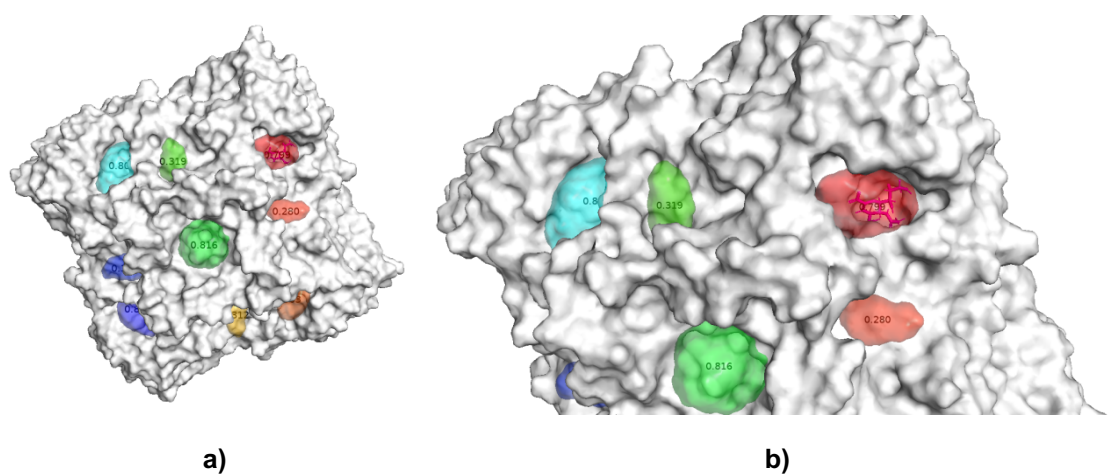

a) b)

**Figure 1 Model predictions for protein 2QWE**. In Figure 1a) the global view of the prediction is shown, where different pockets are depicted in different colors and are labelled with their scores. In Figure 1b) the zoomed of the pocket containing the ligand is shown.

In Figure 1 an example of results of GENEOnet applied to the protein 2QWE is shown. The picture shows a relevant aspect of GENEOnet: the depicted protein is made up of four symmetrical units so that the true pocket is replicated four times. GENEOnet correctly outputs, among the others, four symmetrical pockets which get high scores. This happens thanks to equivariance, because the results of the model on identical units are the same, with position and orientation coherently adjusted. Moreover this happens even if the model has been trained mainly on single chain examples.

**Comparison with other methods**

Making comparisons between different methods of pocket detection is a difficult task as the models can differ in several ways: first of all, despite all the models should start from a file that encodes the structure of a protein, the data that are internally feeding the algorithm can have the most diverse forms. Additionally, different methods optimize different loss functions, which reflect various ways to define the correct detection of a pocket. For this reason usually the results of different models are not immediately comparable.

We saw that many models return as final output a list of pockets ranked with scores, although pockets can be expressed in different ways: as a union of spheres given centers and radii, as a set of dummy atoms given their spatial coordinates or, like in our case, as a set of voxels. Using scores we can perform a comparison based on the ability of the model to assign the highest scores to pockets that match the true ones.

In this way, given our dataset of proteins having only one ligand (see the Methods section for a description of our dataset), and thus one pocket each, we can compute the fraction of proteins whose true pocket is hit by the predicted one with highest score, by the one with second highest score and so on. We say that a predicted pocket A hits the true pocket B if A has the greatest overlap with B. If no predicted pocket has an intersection with the true one, we say that the method failed on that protein. Finally, we compute the cumulative sum of these fractions; in this way we get a curve where the i-th point represents the fraction of proteins whose true pocket has been recognized within the first i highest scored predicted pockets.

In the following we will denote by $H_j$ the proportion of correct recognitions by the j-th top ranked pocket

$$H_j = \frac{\#(\text{proteins whose true pocket is hit by the j} - \text{th top ranked})}{\#(\text{proteins})}$$

and by $T_j$ the corresponding cumulative quantities

$$T_j = \frac{\#(\text{proteins whose true pocket is hit within the j} - \text{th top ranked})}{\#(\text{proteins})} = \sum_{i=1}^{j} H_i$$

In this way different methods can be compared directly: if a model has a cumulative curve that stands above all the others, then that model is definitely better. We chose to use this approach to compare our model with the following other state-of-the-art methods that were available to the authors:

1. Fpocket[38]: a fast geometrical method that employs a detection algorithm based on alpha spheres.

2. P2Rank[43]: a model that uses Random Forests to make predictions on a cloud of points evenly sampled on the Solvent Accessible Surface.

3. DeepPocket[46]: a method that performs a re-scoring of Fpocket cavities by means of 3D CNNs.

4. CAVIAR[31]: a model that uses a novel approach to the classical technique of points enclosure.

5. SiteMap[30]: a model that clusterizes site points based on surface distance and how well they are sheltered from the solvent.

6. CavVis[32]: a model that uses Gaussian surfaces to predict pockets from a visibility criterion.

The comparison has been run on a test set composed by the 75% of our total dataset of molecules, while the remaining 25% has been used as training and validation set (see the Methods section and the Supplementary Information for more details). The results are shown in Tables 1 and 2 and in Figure 2.

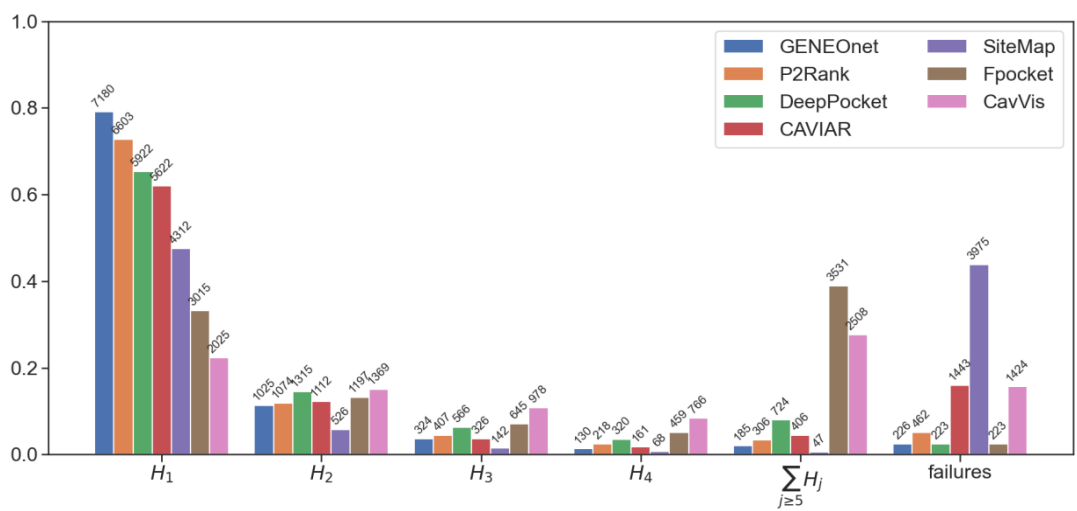

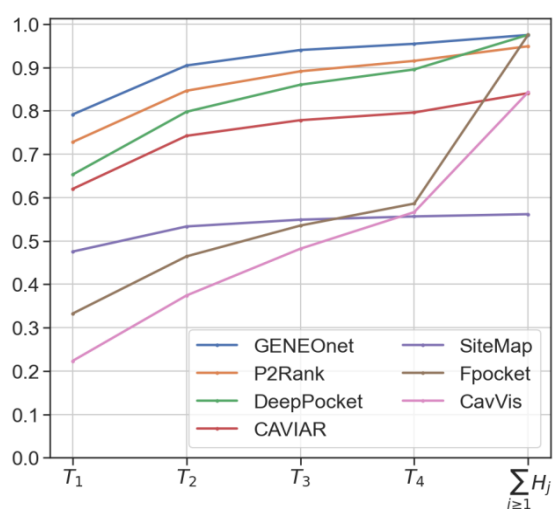

**Figure 2. Results of comparison on the test set.** Figure 2a) shows a bar chart of the $H_j$ coefficients for the different methods, reporting also the absolute frequencies, while Figure 2b) shows cumulative frequencies curves.

GENEOnet has a better performance than all the other methods considered in the comparison; in particular if we look at $T_3$, that is the fraction of proteins whose true pocket is identified within the three top ranked predicted ones, we see that GENEOnet achieves a result of 0.941 which means that, with this data, we can expect that more than 90% of times we will find the right pocket considering just the three pockets with highest score. All the other methods, instead, have a value of $T_3$ below 0.9. We note that SiteMap returned a result just for 7468 proteins out of the total of 12295; proteins for which a result was missing were counted as failures, justifying the high number of failures for this method.

## Discussion

The results obtained on pocket detection confirm many properties that make of GENEOs very promising tools to give rise to efficient techniques of explainable machine learning. In particular we remark that:

- GENEOnet can incorporate prior knowledge: we know for example that lipophilicity, hydrophilicity, etc. , and not only the shape of the molecule, are important in order to detect a binding site, so we introduced a GENEO for each physical or chemical quantity, able to detect areas with the best values for such quantities.

- GENEOnet can forget irrelevant facts: we know that the spatial pose doesn't alter the presence of a pocket, so all our GENEOs are defined to be equivariant with respect to the group of rigid motions of the space.

- GENEOnet needs less parameters: the specific instance of GENEOnet developed for pocket detection has only 17 parameters.

- GENEOnet needs less training examples: we obtained the final model from a training set of just 200 proteins.

- GENEOnet is an interpretable model: after training we can give an understandable meaning to the values of the parameters. For example we know that if a channel has a small convex combination parameter then the corresponding GENEO has little to no effects on the final result.

Moreover GENEOnet provided good results in finding and giving high scores to the most druggable pockets for a given protein. Its performances are better than the other state-of-the-art models considered, since it has significantly lower computational costs and a comparable or, in many cases, even better accuracy than the other methods.

We have anyway to remark that GENEOnet is a quite powerful tool in situations in which some prior knowledge on the problem is available. When such knowledge is not available, traditional methods of deep or machine learning are surely more effective than GENEOnet. Additionally the phase of design of the right GENEOs, able to incorporate the previous knowledge on the problem, can be long and rather complicated. We are anyway working to setup possible "libraries" of GENEOs, able to detect specific equivariance properties, that may be used and combined also by non-experts to define new architectures for new problems.

Furthermore, some aspects of GENEOnet applied to pocket detection could still be improved. For example, we could introduce GENEOs that can work on more than one channel simultaneously; the coefficient `k` of the accuracy function (see the Methods section) has been fixed, while it could be optimized together with the other parameters; other convolutional kernels, possibly depending on more than one parameter each, could be tested; other optimization methods to maximize the accuracy function could be used. We will work in these directions, to further improve GENEOnet for protein pocket detection.

By the way, the quite promising results obtained for our specific application make us believe that GENEOs may pose the basis for a new mathematical foundation, to develop techniques of explainable deep learning and new kinds of neural networks. This and many other studies confirm the importance of exploiting prior geometrical knowledge on the features of data. We believe that GENEOs are, at the moment, one of the best and more mathematically sound way to do that, even if there are still many aspects to deepen both on a theoretical and experimental point of view. We plan to bring ahead research on the theory of GENEOs to strengthen the mathematical core of our approach and to find other successful applications of GENEOnet to reinforce the experimental evidence of its capabilities.

## Methods

### GENEO definition

Let us formalize the concept of group-equivariant non-expansive operator.

We assume that a space $\Phi$ of functions from a set $X$ to $\mathbb{R}^k$ is given, together with a group $G$ of permutations of $X$, such that if $\varphi \in \Phi$ and $g \in G$ then $\varphi \circ g \in \Phi$. We say that $(\Phi, G)$ is a *perception pair*.

Let us assume that another perception pair $(\Psi, H)$ is given, and fix a homomorphism $T: G \to H$. A map $F: \Phi \to \Psi$ is called a *group equivariant non-expansive operator (GENEO)* if the following conditions hold:

1. $F(\varphi \circ g) = F(\varphi) \circ T(g)$ for every $\varphi \in \Phi$, $g \in G$ (equivariance)

2. $||F(\varphi_1) - F(\varphi_2)||_\infty \leq ||\varphi_1 - \varphi_2||_\infty$ for every $\varphi_1, \varphi_2 \in \Phi$ (non-expansiveness).

**Data**

Protein-ligand complexes were retrieved from the PDBbind database v.2020[48]. This database provides a comprehensive and curated collection of experimentally measured binding affinity data for protein-ligand complexes deposited in the Protein Data Bank (PDB)[49], which is the largest resource for experimentally determined biomolecular structures. The aim of the PDBbind database is to provide high-quality datasets for the development of structure-based drug design methods.

Protein structures were preprocessed using Schrodinger Protein Preparation Wizard utility[50], with default settings. A total of 12295 protein-ligand complexes have been retained for the experiments.

In order to automate the input preparation for GENEOnet, the GENEOprep software was developed, written in C++ and based on VEGA, a molecular modelling suite of programs developed in our laboratories[51]. In particular, GENEOPrep 1) reads the PDB file, 2) adds the hydrogens if they are missing, 3) assigns the atomic charges according to the Gasteiger-Marsili method[52], 4) assigns the Broto-Moreau lipophilicity parameters[53], 5) searches for atoms which are H-bond acceptors/donors 6) assigns the non-bond parameters according to CHARMM 36 force field[54], and finally 7) saves the atom coordinates and the computed data to a CSV file that is used to compute the potentials described hereafter.

Input data have been modelled as bounded functions from the Euclidean space $\mathbb{R}^3$ to $\mathbb{R}^d$, in this specific case we chose $d = 8$, that is the number of distinct geometrical, chemical and physical potential fields that we computed and took into account for the analysis. We selected the potentials reported in Table 3, in the following we will also call these functions *channels*, borrowing a typical computer vision terminology.

Eventually we need to define the ground truth that will be used later to assess the accuracy of predictions: we chose to identify a "pocket" with the spatial region occupied by the ligand.

Actually proteins can have multiple ligands, however the PDBbind contains only complexes where proteins are coupled with their most representative ligand (this choice was done by the scientist that first studied the protein, usually the ligand that shows the highest biological activity with the protein is chosen), moreover for each protein we selected only those chains that "touched" the ligand. In this way we avoided repetitions of the same cavity due to symmetries of the protein and we got a set of molecules hosting only one ligand, and thus defining only one true pocket.

**The model**

The input data are fed to a first layer of GENEOs chosen from a set of parametric families of operators, each one defined by a shape parameter $\sigma_i$. These families were designed to include the a priori knowledge of the experts of medicinal chemistry.

We opted for convolutional operators whose properties can be completely determined by the nature of their kernels. Moreover by making the i-th kernel dependent only on one shape parameter $\sigma_i$, we have direct control on the action of each operator.

We used Gaussian kernels for all channels but the Distance and the Electrostatic field since, except for these two fields, we look for regions with homogeneous values inside a pocket. For the Distance field we defined a spherical kernel using the radial function $K_1(r) = \frac{1}{2} - \tanh(h \cdot (r - \sigma_1)) + \frac{1}{2}\tanh(h \cdot (r - (\sigma_1 + \beta)))$, where $r$ is the distance from the origin. This is a differentiable approximation of the step function $\widetilde{K_1}(r) = \mathbb{I}_{\{r \leq \sigma_1\}}(r) - \mathbb{I}_{\{\sigma_1 < r \leq \sigma_1 + \beta\}}(r)$ which takes value +1 inside a sphere of radius $\sigma_1$ centered at the origin, -1 inside the spherical shell from radius $\sigma_1$ to $\sigma_1 + \beta$ and 0 outside the sphere of radius $\sigma_1 + \beta$. The negative value inside the spherical shell ensures that positive regions far from the protein surface will not have a high response after the convolution. In this way the kernel will detect only spherical voids close to the protein that are surrounded by protein atoms. The parameter $\beta \in \mathbb{R}_+$ tunes how much we are interested in the region outside the inner sphere: higher values mean that we look for pockets buried deep inside the protein. Finally the parameter $h \in \mathbb{R}_+$ measures how $K_1$ is a good approximation of $\widetilde{K_1}$, where higher values lead to a better approximation, but to almost non-differentiability, that could be an issue during the optimization step.

For the Electrostatic field we used a Gaussian Laplacian kernel as we are interested in finding variations in the charges of atoms inside the pocket. Except for the kernel of the Distance field, where the shape parameter $\sigma_1$ is the radius of the positive sphere, for all the other kernels the shape parameters are the standard deviations of the Gaussians.

Despite being different, these operators share a common feature: their kernels are defined through rotationally invariant functions. This fact, together with the properties of convolution, guarantees that the corresponding GENEOs satisfy the key requirement to be equivariant with respect to the group of isometries of $\mathbb{R}^3$.

In the second step these operators are combined through a convex combination, with weights $\alpha_1, \ldots, \alpha_8$ in order to obtain a composite operator $F_\alpha(\cdot) = \sum_{i=1}^{d} \alpha_i F_i(\cdot)$ whose output is the normalized function $\psi$ from $\mathbb{R}^3$ to $[0,1]$; here $\psi(x)$ can be read as the probability that a point $x \in \mathbb{R}^3$ belongs to a pocket. Finally, given a probability threshold $\theta \in [0,1]$, we get the different pockets returned by the model by taking the connected components of the superlevel set $\{\psi \geq \theta\} \subseteq \mathbb{R}^3$.

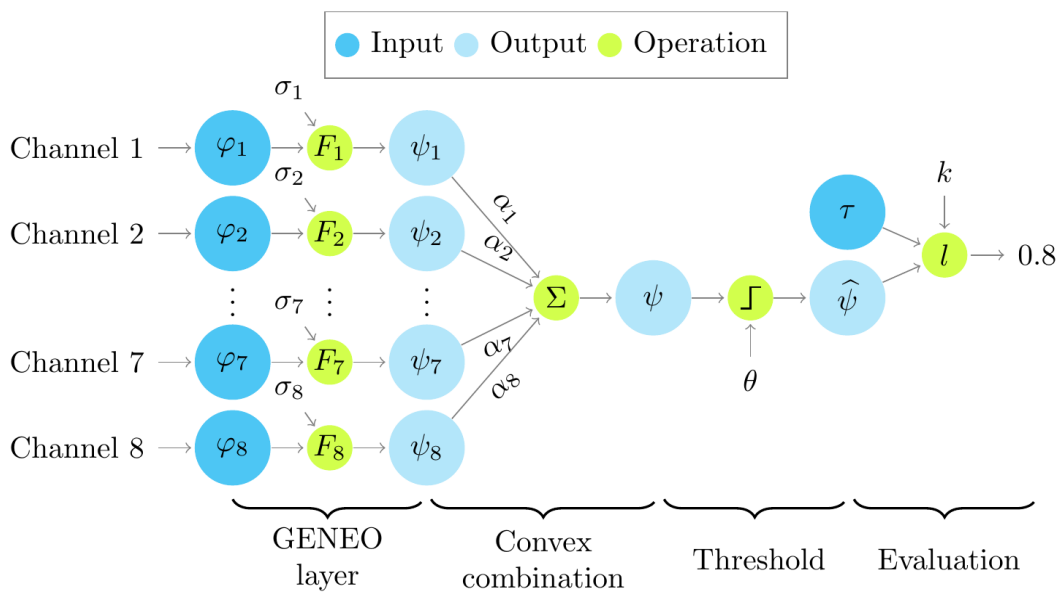

**Figure 3 Model workflow.** The input channels $\varphi_1, \ldots, \varphi_8$ are fed to the GENEOs $F_1, \ldots, F_8$ that depend on the shape parameters $\sigma_1, \ldots, \sigma_8$, this first layer returns the intermediate outputs $\psi_1, \ldots, \psi_8$. These outputs are combined through convex combination with weights $\alpha_1, \ldots, \alpha_8$ to get the final result $\psi$. To obtain pockets a thresholding operation with a parameter $\theta$ is applied to $\psi$, producing the binary function $\widehat{\psi}$, which finally can be compared to the ground truth $\tau$ through the accuracy function.

In order to implement the method, we discretized the space around the protein using a cubical grid of voxels, having side $\delta = 1.25$Å in all directions and using a bounding box slightly bigger than the size of the molecule, to avoid problems with pockets very close to the border. In each voxel we computed constant approximations of the input channels, producing thus a spatial discretization of the considered potential fields. A similar discretization has been applied to the GENEOs, that have been represented as discrete convolutional operators, with kernels formed by three-dimensional cubical arrays of size 17. The choice of convolutional operators at this point allowed us to exploit the efficient implementation of discrete convolution.

**Parameter identification**

The model that was described so far, as shown in Figure 3, has a total of 17 parameters ($\{\sigma_i\}_{i=1}^{8}$, $\{\alpha_j\}_{j=1}^{8}$ and $\theta$). The codes were written using both C and Python. The fact that the model only employs convolutional operators, and their linear combinations, allowed us to set up an optimization pipeline quite similar to the one of a 3D Convolutional Neural Network (CNN), but with two fundamental differences. First of all our model has a really tiny set of parameters, if compared to a classical CNN: we estimated that a recent method called DeepSite[45], which implements a classical 3D CNN for pocket detection, has 844529 parameters; DeepPocket[46], an even newer approach that uses a 3D CNN to rescore Fpocket[38] predictions, has 665122 parameters. Additionally the convolutional kernels of the GENEOs are not learned entry by entry as in classical CNNs, since in this way equivariance would not be preserved at each iteration; instead at each step the kernels are recomputed from the shape parameters that are updated during the optimization.

To identify the unknown parameters, we have to optimize a cost function that evaluates the goodness of our predictions. If we call $\hat{\psi}$ the output of the model after thresholding then we must compare it to the ground truth represented by the binary function $\tau$, that takes value 1 in those voxels occupied by the ligand and 0 in the other voxels. We adopted the following accuracy function that needs to be maximized:

$$\ell(\hat{\psi}, \tau) = \frac{|\hat{\psi} \wedge \tau| + k |(\mathbf{1} - \hat{\psi}) \wedge (\mathbf{1} - \tau)|}{|\tau| + k |\mathbf{1} - \tau|} \in [0,1]$$

With $|\hat{\psi} \wedge \tau|$ we denote the minimum between the two functions, with $|\cdot|$ the volume of the set where the function equals 1 and with $\mathbf{1}$ a constant function equal to 1.

Note that this definition of $\ell(\hat{\psi}, \tau)$ is well posed since all our functions are defined only on a (voxelized) compact cubic region surrounding the molecule.

The hyperparameter $k$ ranges in $[0,1]$, and when $k = 1$ the accuracy function is simply the fraction of correctly labelled voxels out of the total.

We note that assuming $k = 1$ would give more importance to correctly recognize the voxels which do not host the ligand (that is the "non-cavity"), than to correctly recognize the voxels where the ligand is located (that is the cavity), because the ligand, and thus the cavity, has usually a much smaller size than the total volume of the protein. Thus assuming $k = 1$ would not focus the method on the research of the true pocket and, moreover, we would not be able to identify other possible pockets that could be good candidates to host other ligands.

For these reasons we choose $k < 1$, which allows to balance the two terms of the sum in the numerator to obtain more and slightly bigger pockets.

The other extreme situation happens when $k = 0$. In this case the model generates pockets that can be as big as the whole grid since the only thing that matters is that every voxel containing the ligand is marked with 1. Keeping this in mind, the choice of a $k$ strictly between 0 and 1 gives the best results. In particular we found that values of $k$ in the interval $[0.01, 0.05]$ give similar results, all characterized by a rather small number of pockets of suitable size.

**Pockets ranking**

Eventually, pockets are found as the connected components of the thresholded output of the model. In this way we get an array where voxels located in a pocket are labelled with the progressive number of the connected component they belong to, while they are labelled with 0 if they do not belong to a pocket. Actually this representation is not much informative, since in the applications of pockets detection in medicinal chemistry it is desirable to compute also the *druggability* of the identified cavities, that is a ranking of the pockets on the basis of their fitness to host a ligand.

To give a score to each pocket, we went back to the output of the model before the thresholding, that is to the function $\psi(x)$ which was interpreted as the probability that a voxel $x$ belongs to a pocket. The score of a pocket was then computed as the average value of $\psi$ computed only on the voxels belonging to the pocket, rescaled by a factor which takes into account the volume of the pocket, in order to avoid giving too high scores to very small pockets.

Eventually the final output of the model consists in a list of pockets, given as the coordinates of their voxels, and the corresponding scores.

**Training**

The optimization of $\ell(\hat{\psi}, \tau)$ was performed using the Adam optimizer[55], with different learning rates for the three groups of parameters (the $\sigma_i$'s, the $\alpha_j$'s and $\theta$) that are involved in the pipeline of Figure 3. We used this strategy since the shape parameters tend to have smaller gradients and so they benefit from a slightly higher learning rate.

As training set we selected a random set of 200 proteins, since from empirical evidence (see the Supplementary Information) increasing further the size of the training set does not lead to significant changes in parameter estimates. In order to study the robustness of the estimation procedure, and to optimize, not only the recognition of the cavity hosting the ligand, but also its ranking in the pockets identified on each molecule, the sampling of the training set and the optimization was repeated 200 times, obtaining thus 200 models based on different training sets of the same size. The optimization of each of the 200 models started from the same initial guess for the parameters. A validation set composed by 3073 molecules (that is about the 25% of the entire dataset) was used to compute the $H_j$ coefficients of the 200 models, and the model maximizing $H_1$ was selected as the best.

The optimization was accelerated by computing and storing at the beginning the values of the potentials used for each channel on the molecules of the training set. In this way the training time for 50 epochs of the optimization algorithm, is ~6' with GPU acceleration and ~40' without.

In Table 4 the estimated values of the parameters for the best model are reported. Their values can be used to provide further interpretation to the results: the relative magnitude of the shape parameters $\sigma_i$ reflect their influence on the convolution kernels, while the values of the convex combination parameters $\alpha_j$ can be regarded as the corresponding channel's importance. So for example we see from Table 4 that the operator working on the electrostatic field has a small combination parameter $\alpha_3 = 0.054$, meaning that the result of this specific GENEO is not very important for the final result. The highest weights are $\alpha_1 = 0.362$ and $\alpha_4 = 0.338$ that correspond to the operators working on the distance and lipophilic field, meaning that the geometry and the lipophilic properties are the most significant for the recognition of pockets. These interpretations are also supported by a simulated study of the distribution of the parameters which is reported in the Supplementary Information.

## Acknowledgements

Funding from Dompè Farmaceutici S.p.A. to run this project is acknowledged by the authors. Additionally P.F. has been partially supported by INdAM-GNSAGA.

## Author contributions statement

G.B., P.F., A.M., A.P. conceived the idea and wrote the paper; G.B. wrote the codes; F.L., A.P., C.G. preprocessed the data; G.B., F.L., C.G. run the experiments; A.R.B., C.T. supervised the experiments.

## Additional information

**Competing interests** A.R.B., F.L. and C.T. are employees of Dompè Farmaceutici S.p.A..

## Data Availability

No datasets were generated during the current study. The analyzed molecules are available at https://www.rcsb.org/

## Code Availability

The code used to generate the results shown in this study will be property of Dompè Farmaceutici S.p.A. and will be available under request.

## Tables

| Method | $H_1$ | $H_2$ | $H_3$ | $H_4$ | $\sum_{j\geq 5} H_j$ | failures |
|---|---|---|---|---|---|---|
| GENEOnet | 0.792 | 0.113 | 0.036 | 0.014 | 0.020 | 0.025 |
| P2Rank | 0.728 | 0.118 | 0.045 | 0.024 | 0.034 | 0.051 |
| DeepPocket | 0.653 | 0.145 | 0.062 | 0.035 | 0.080 | 0.025 |
| CAVIAR | 0.620 | 0.123 | 0.036 | 0.018 | 0.044 | 0.159 |
| SiteMap | 0.475 | 0.058 | 0.016 | 0.007 | 0.006 | 0.438 |
| Fpocket | 0.332 | 0.132 | 0.071 | 0.051 | 0.389 | 0.025 |
| CavVis | 0.223 | 0.151 | 0.108 | 0.084 | 0.277 | 0.157 |

**Table 1 Fractions of correct recognition.**

| Method | $T_1$ | $T_2$ | $T_3$ | $T_4$ | $\sum_{j\geq 1} H_j$ |
|---|---|---|---|---|---|
| GENEOnet | **0.792** | **0.905** | **0.941** | **0.955** | **0.975** |
| P2Rank | 0.728 | 0.846 | 0.891 | 0.915 | 0.949 |
| DeepPocket | 0.653 | 0.798 | 0.860 | 0.895 | **0.975** |
| CAVIAR | 0.620 | 0.743 | 0.779 | 0.797 | 0.841 |
| SiteMap | 0.475 | 0.533 | 0.549 | 0.556 | 0.562 |
| Fpocket | 0.332 | 0.464 | 0.535 | 0.586 | **0.975** |
| CavVis | 0.223 | 0.374 | 0.482 | 0.566 | 0.843 |

**Table 2 Cumulative fractions of correct recognition**

| Name | Type | Expression | Notes |
|---|---|---|---|
| Distance | Geometrical | $\varphi_1(x) = d(x, x_{a^*}) - r_{a^*}$ | $x_{a^*}$ and $r_{a^*}$ are coordinates and radius of the nearest atom to the point $x$. |
| Gravitational | Geometrical | $\varphi_2(x) = \sum_{a \in A} \frac{m(a)}{d(x, x_a)}$ | $m(a)$ is the mass of atom $a$. |
| Electrostatic | Physical | $\varphi_3(x) = \sum_{a \in A} \frac{q(a)}{d(x,x_a)}$ | $q(a)$ is the partial charge of atom $a$. |
| Lipophilic | Chemical | $\varphi_4(x) = \sum_{a \in A} \frac{l(a)}{d(x,x_a)}$ | $l(a)$ is the lipophilic coefficient of the atom $a$ if it is negative, 0 otherwise. |
| Hydrophilic | Chemical | $\phi_5(x) = \sum_{a \in A} \frac{h(a)}{d(x,x_a)}$ | $h(a)$ is the lipophilic coefficient of the atom $a$ if it is positive, 0 otherwise. |
| Polar | Chemical | $\phi_6(x) = \sum_{a \in A} \frac{p(a)}{d(x,x_a)}$ | $p(a)$ is 1 if atom $a$ is polar, 0 otherwise. |
| HB Acceptor | Chemical | $\phi_7(x) = \sum_{a \in A} -\varepsilon_a (R^6 - 2R^4)$ | $R = R_{min}/d(x,x_a) + 0.96$ where $\varepsilon_a$ and $R_{min}$ are parameters of the specific type of atom. |
| HB Donor | Chemical | $\phi_8(x) = \sum_{a \in A} -\varepsilon_a (R^6 - 2R^4) \cos^2 \varphi_1 \cos^2 \varphi_2$ | $R = R_{min}/d(x,x_a) + 0.96$ where $\varepsilon_a$ and $R_{min}$ are parameters of the specific type of atom, $\varphi_1$ and $\varphi_2$ are angles defined by triples of points involved in the bond. |

**Table 3 Channels definition.** List of potentials that have been used to build GENEOnet. In some cases, constants have been ignored because of the subsequent normalization. A summation over all the atoms of the protein would be unfeasible, but, since many potentials depend on the inverse of the distance from $x$, in our computations we neglected atoms too far apart from $x$ and the above sums have been computed only for $a \in N(x)$, where $N(x)$ is a suitable neighbourhood of $x$.

| Unit | Channel | $\sigma_i$ | $\alpha_j$ | $\theta$ |
|---|---|---|---|---|
| 1 | Distance | 3.110 | 0.362 | 0.756 |
| 2 | Gravitational | 5.197 | 0.002 | |
| 3 | Electrostatic | 2.561 | 0.054 | |
| 4 | Lipophilic | 4.678 | 0.338 | |
| 5 | Hydrophilic | 3.545 | 0.001 | |
| 6 | Polar | 6.166 | 0.185 | |
| 7 | HB Acceptor | 4.186 | 0.056 | |
| 8 | HB Donor | 3.908 | 0.001 | |

**Table 4 Parameters of the optimized model**.

# GENEOnet: A new machine learning paradigm based on Group Equivariant Non-Expansive Operators. An application to protein pocket detection. Supplementary Information.

## Size of the training set

In order to choose the appropriate size for the training set we made the following experiment: we iteratively trained a family of 31 models starting from a training set of size 5 up to a training set of size 305. At each step a new model is obtained starting from the same initial guess for the parameters, but with a training set whose size is augmented by 10. The added molecules are randomly sampled in the entire database, as well as the starting 5. For each iteration we recorded the values of the parameters after the optimization and the results are shown in Figure 1.

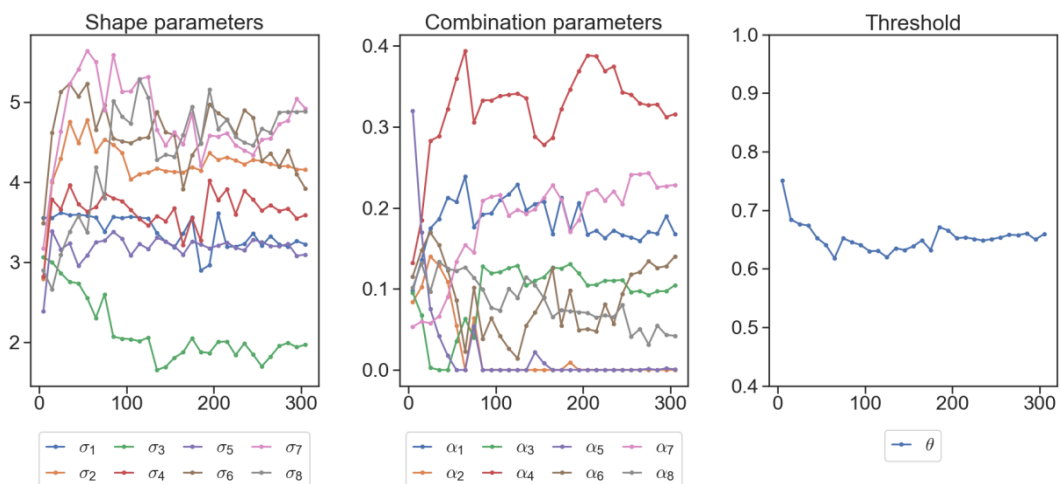

**Figure** Errore. Solo documento principale.**: Parameters evolution** This Figure shows the evolution of the values of the optimal parameters as function of the size of the training set. The leftmost plot is relative to the shape parameters, the central one to the convex combination parameters and the rightmost one to the threshold.

The plots show a progressive stabilization of most of the parameter values as the size of the training set increases. This behavior could be expected, since the small number of parameters of GENEOnet does not need a huge training set to be identified. Consequently we chose to adopt a training set of size 200, since from the plots we can see that almost all the parameters have reached a stable value (still accepting a little fluctuation) and this choice proves to be a good compromise between accuracy and training times reduction.

## Choice of the training set and estimation robustness

We generated 200 different models all starting from the same initial values of the parameters but with different training sets of size 200. Each training set was randomly sampled and it was used to train the corresponding model. The optimal values of the parameters were recorded and used to generate the box plots in Figure 2.

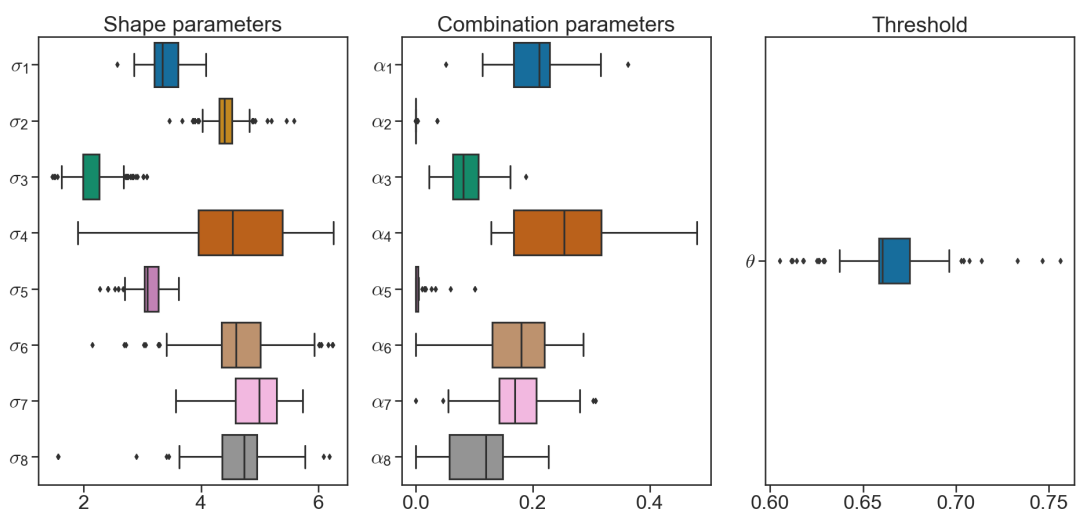

**Figure** Errore. Solo documento principale.**: Optimal parameter distributions** These box plots show the distributions of the parameters. In the leftmost panel the distributions of the shape parameters are plotted, in the central panel those of the convex combination parameters are reported and in the right panel the distribution of the threshold is depicted.

The boxplots may help to provide deeper interpretations of the parameters: for example, the overall distribution of $\alpha_5$, that is the coefficient of the Hydrophilic channel, is very concentrated towards 0 meaning that the corresponding estimated coefficient will rarely be much higher than 0; consequently the Hydrophilic channel will always have a low importance for the final pocket prediction. On the contrary, the distributions of $\alpha_1$ and $\alpha_4$ are spread around values quite above 0, meaning that the Distance and the Lipophilic channels, of which $\alpha_1$ and $\alpha_4$ are the coefficients, respectively, seem to always give a significant contribution to the final prediction.

The 200 trained models are optimal with respect to the accuracy function $\ell(\hat{\psi}, \tau)$ defined in the Methods section, but this does not guarantee that the models are optimal also with respect to the scoring of the pockets. In order to select the overall best model, we compared them on the same validation set of size 3073 (almost 25% of the total 12295 proteins) by computing the coefficients $H_j$ and ordering the models for decreasing values of $H_1$. We note that all the training sets have a small random intersection with the validation set: the mean size of this intersection is 52, with a minimum of 35 and a maximum of 64 as we could expect since the training sets are sampled uniformly at random from the entire dataset. Anyway, this fact is negligible, since the intersection is always smaller than the 2% of the validation set.

The results are shown in Table 1.

| index | $H_1$ | $H_2$ | $H_3$ | $H_4$ | $\sum_{j\geq 5} H_j$ | failures |
|---|---|---|---|---|---|---|
| 148 | 0,805 | 0,102 | 0,033 | 0,014 | 0,017 | 0,028 |
| 116 | 0,789 | 0,107 | 0,039 | 0,015 | 0,024 | 0,027 |
| 18 | 0,787 | 0,112 | 0,036 | 0,017 | 0,021 | 0,028 |
| 62 | 0,786 | 0,113 | 0,044 | 0,014 | 0,021 | 0,022 |
| 115 | 0,786 | 0,113 | 0,044 | 0,014 | 0,021 | 0,022 |
| 88 | 0,785 | 0,118 | 0,038 | 0,015 | 0,023 | 0,021 |
| 150 | 0,782 | 0,116 | 0,038 | 0,013 | 0,024 | 0,027 |
| 67 | 0,782 | 0,116 | 0,038 | 0,013 | 0,024 | 0,027 |
| 181 | 0,781 | 0,115 | 0,044 | 0,013 | 0,019 | 0,029 |
| 57 | 0,781 | 0,121 | 0,038 | 0,015 | 0,022 | 0,023 |

**Table** Errore. Solo documento principale.**: Performances of the best models according to the validation set** We report the performances of the ten models with highest $H_1$ as computed on the validation set.

Table 1 shows on top the model that was chosen as our final version of GENEOnet, and its training set was considered optimal both for the accuracy function $\ell(\hat{\psi}, \tau)$ and for the scoring of the pockets. This model was used to perform the comparison described in the Results section. We remark that while we admitted a small intersection between the training and the validation set, we avoided any overlapping between the training set of the best selected model and the test set, in order to obtain unbiased results in the comparison with the other methods for pocket detection. Eventually the sizes of the actual employed sets are:

1. 200 for the chosen training set (approximately 1.5% of the whole dataset)
2. 3073 for the validation set (approximately 25% of the whole dataset) with 48 proteins in common with the chosen training set.
3. 9070 for the test set (approximately the 74% of the whole dataset) that is totally disjoint from the training and the validation sets, after removal of the common proteins.